# A Deep Learning Model for Atomic Structures Prediction Using X-ray Absorption Spectroscopic Data

Liang Li,[1,a)] Mindren Lu,[1,2] and Maria K. Y. Chan[1,b)]

[1]Center for Nanoscale Materials, Argonne National Laboratory, Lemont, Illinois 60439, USA

[2]Department of Electrical Engineering and Computer Science, Massachusetts Institute of Technology,

Cambridge, Massachusetts 02139, USA

A deep neural network (DNN) model consisting of two hidden layers was proposed for predicting the immediate environments of specific atoms based on X-ray absorption near-edge spectra (XANES). The output layer of the DNN can be adjusted to form a classifier or regressor, to predict the local and overall coordination environments, respectively. Using $Li_3FeO_{3.5}$ as a model system, it was demonstrated that the prediction accuracy of the DNN classifier is higher than 98%, and the predictions of the DNN regressor also showed notable agreement with the ground truth. Therefore, despite its simplicity, this DNN architecture can be expected to be generally capable of predicting the structural properties of various systems. Fine tuning of the hyperparameters, bias-variance tradeoff, and strategies to enrich the versatility of the model were also discussed.

Due to the high sensitivity to local structures and electronic configurations, X-ray absorption near-edge spectroscopy (XANES) has been an indispensable tool in a wide range of applications, including measuring the electronic structures of compounds,[1] monitoring site-specific reactions in large molecules,[2] and probing the immediate coordination environment and oxidation states of catalytic and electrochemical systems.[3,4] Experimentally, deciphering the underlying atomistic structures through XANES is primarily performed by correlating the measured spectra with previously-measured reference spectra of standard materials, for which the structures and compositions have already been properly resolved. However, aside from the inherent laboriousness associate with this procedure, such a correlation heavily hinges upon the choice and purity of the reference materials, and is thus inevitably prone to biased interpretations, not to mention the added time and expense of measuring reference spectra for each experiment. Extensive effort has been devoted to establishing systematic routines to correlate XANES and material structures, and with the developments in theories and computational techniques, several theoretical methodologies

[a)]Electronic mail: lli14@binghamton.edu
[b)]Electronic mail: mchan@anl.gov

have proven effective and reliable in predicting XANES that match experimental observations well, once the underlying atomic configurations are established.[5–7] On the other hand, the inverse problem, i.e., inferring the structures and chemical properties of materials based on the observed spectra, has also attracted considerable attention in recent years, owing to the development in machine learning techniques and the advancement of theoretical tools that enable the generation of sufficient data for training.[8–10] In particular, deep neural networks (DNN) have shown great promise in capturing the nonlinear relationships between spectral features and the underlying atomistic structures,[9,10] and through proper calibration, DNN generally has great generalization performance, which refers to the ability to accurately predict the labels of previously unseen data in the supervised learning framework.

In this letter, a simple DNN architecture is presented that can predict the local coordination environments of specific atoms and overall stoichiometry of compounds. The DNN was trained on oxygen K-edge XANES of $Li_3FeO_{3.5}$, an intermediate composition formed during the electrochemical delithiation of a Li-ion battery material, $Li_5FeO_4$.[4,11] $Li_3FeO_{3.5}$ is characterized by a disordered structure with a variety of oxygen local coordination environments, and precise identification of these environments at the atomistic scale is of great importance in understanding its physical and electrochemical properties. The number of Fe neighbors surrounding individual oxygen atoms is an effective indicator of the degree of covalency between Fe and oxygen, and also dictates the nature of the charge compensation mechanism during electrochemical reactions.[11,12] As shown in Fig. 1, the Fe coordination number of individual oxygen atoms ranges from 0 to 4 (Figs. 1(a-e)). Oxygen dimers, which greatly influence the capacity and stability of the battery materials, and are deduced to exist as intermediates due to the observation of oxygen evolution,[11] were also found in *ab-initio* molecular dynamic (AIMD) simulations of $Li_3FeO_{3.5}$ under elevated temperatures (Fig. 1(f)).[4] Fig. 1 presents corresponding oxygen K-edge spectra of the abovementioned structures, computed using the *OCEAN* package that implements the Bethe-Salpeter Equation (BSE).[13] Some spectra display distinct features in terms of relative peak intensities and peak positions, but others exhibit subtle differences that are difficult to distinguish. $Li_3FeO_{3.5}$ thus serves as a good model system to examine the capability of the DNN in predicting oxygen atom local environments based on the spectral features. While the computed spectra for individual atoms provide a testbed for the DNN model, realistically, the measured XANES contain collective information of multiple oxygen atoms under various local environments, and a more relevant problem would be predicting the sample-averaged Fe coordination number based on the superposition of an ensemble of spectra, a problem which can also be tackled by using a DNN.

The identification of oxygen local environments, or equivalently, prediction of Fe coordination numbers, is essentially a multi-class classification task, since the local environments can be described by one of the classes listed in Fig. 1, depending on the number of coordinated Fe atoms and whether an O-O bond is formed, and different classes are mutually exclusive. In contrast to the discrete outputs in the classification task, for the prediction of the average Fe coordination numbers over an ensemble of oxygen atoms, a regression model is necessary to sufficiently reproduce the continuous quantitative outputs. Therefore, a DNN classifier and a DNN regressor were constructed for predicting the Fe coordination numbers of individual, and a collection of, oxygen atoms, respectively.

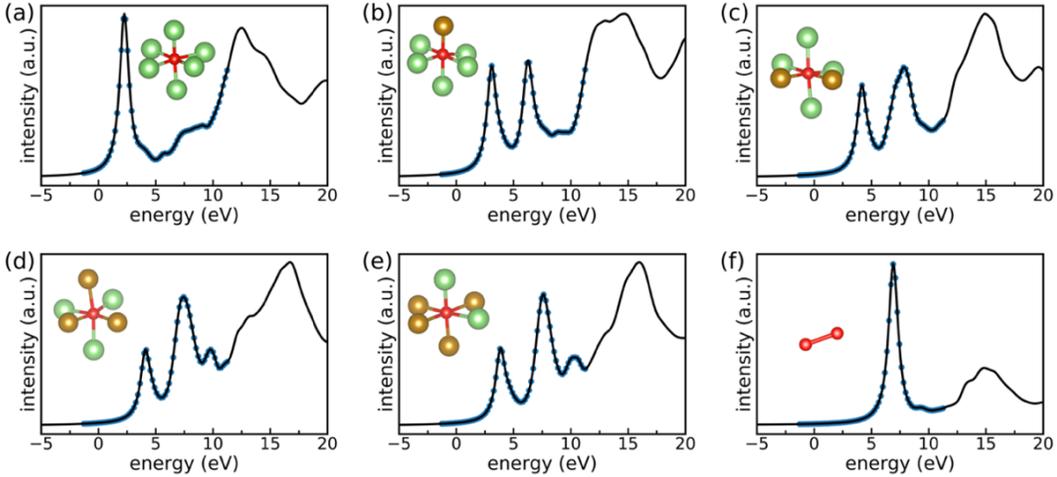

FIG. 1. Typical computed K-edge XANES of oxygen atoms with various local environments in $Li_3FeO_{3.5}$: oxygen atom coordinated by (a) zero (b) one (c) two (d) three and (e) four Fe atoms, (f) oxygen dimers. Due to the incapability of computing the absolute photon energies within the pseudopotential framework, the zero energies denote the conduction band minimum of the ground-state systems with no excitons. Atomic models of corresponding local structures are shown in the insets, with green, brown and red spheres representing Li, Fe and O atoms, respectively. Blue dots indicate the data points used in DNN training.

AIMD simulations were first performed to model the structural change of $Li_3FeO_{3.5}$ under 300 K and 1000 K, respectively. Detailed AIMD procedure was outlined elsewhere.[4] Twenty-one snapshots were sampled along the AIMD trajectory, and the O K-edge XANES were computed using the *OCEAN* package.[13] Each $Li_3FeO_{3.5}$ simulation cell contains 49 O atoms, and approximately 1000 raw spectra were obtained from the XANES modeling. To mitigate the high computational cost of BSE and generate enough spectral data to train and calibrate the DNNs, synthetic training examples for the DNN classifier were constructed by averaging existing raw spectra that belong to the same classes. For the DNN regressor, to mimic the spectra collected from a compound with multiple oxygen atoms, synthetic spectra were formed by averaging $n$ raw spectra, where $n$ ranges from 20 to 70. Corresponding training labels were then obtained by averaging the Fe coordination numbers of the $n$ individual spectra used for synthetic data generation. This is similar to the procedure described in a previous work.[9] For all spectra, 70 data points were uniformly sampled in the near-edge region that spans between -1.2 and 11.5 eV, and these sampled intensity values, as highlighted in Fig. 1, were used as the inputs for DNNs.

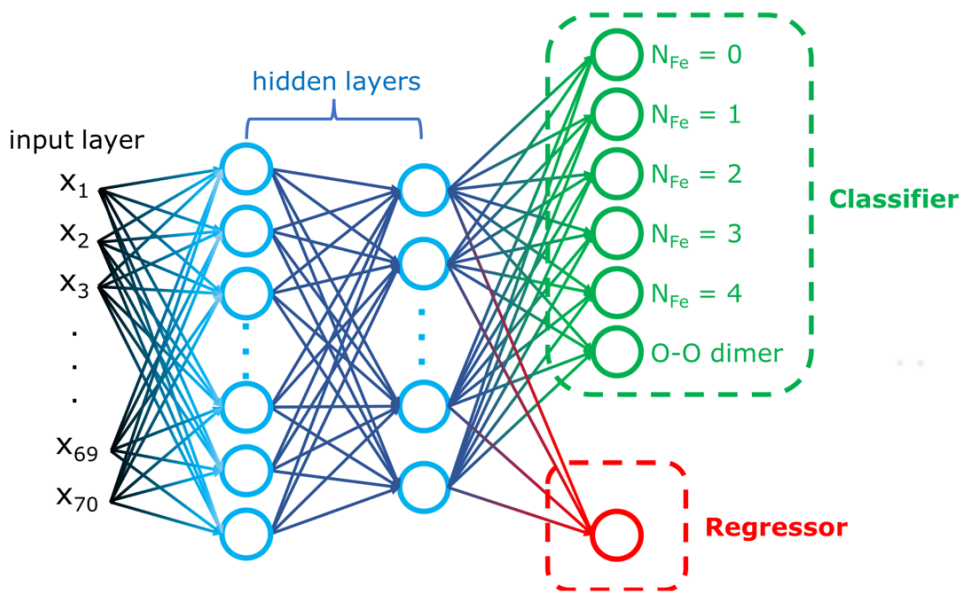

FIG. 2. Schematic of the DNN architectures. Both the DNN classifier and regressor are composed of two hidden layers and one output layer. The output layer of the classifier has six nodes, whereas the regressor output layer has one node.

The DNNs were trained using TensorFlow.[14] The entire datasets used for the DNN classifier and regressor training were both divided into training, validation, and test sets, with a ratio of 3:1:1. Fig. 2 shows the DNN architectures proposed in this work. For both the multi-class classification and the regression problems, a three-layer DNN (two hidden layers and one output layer) was found to yield reasonable prediction accuracies on the validation and test sets. For the DNN classifier, the output layer contains 6 nodes, corresponding to the 6 classes of oxygen local environments as discussed earlier. The softmax activation function was used in the output layer to calculate the categorical probability distribution, and the cross-entropy loss was used as the loss function. For the DNN regressor, the output layer contains a single node, with no activation function being applied. The mean squared error (MSE) between the predicted average Fe coordination numbers and the training labels served as the loss function. Both networks utilized the rectified linear unit (ReLU) as the activation function for the hidden layer nodes, and the first and second hidden layers contain 8 and 6 nodes, respectively, which was found to result in the optimal performance, as discussed later. Xavier initialization[15] was used to initialize the weights in all layers. During training, *L2* regularization was applied to the weights in each layer to prevent overfitting, with a regularization parameter of 0.1. The Adaptive Moment Estimation (Adam) method[16] was adopted to update the weights to minimize the loss functions. Mini-batch gradient descent was also applied during training.

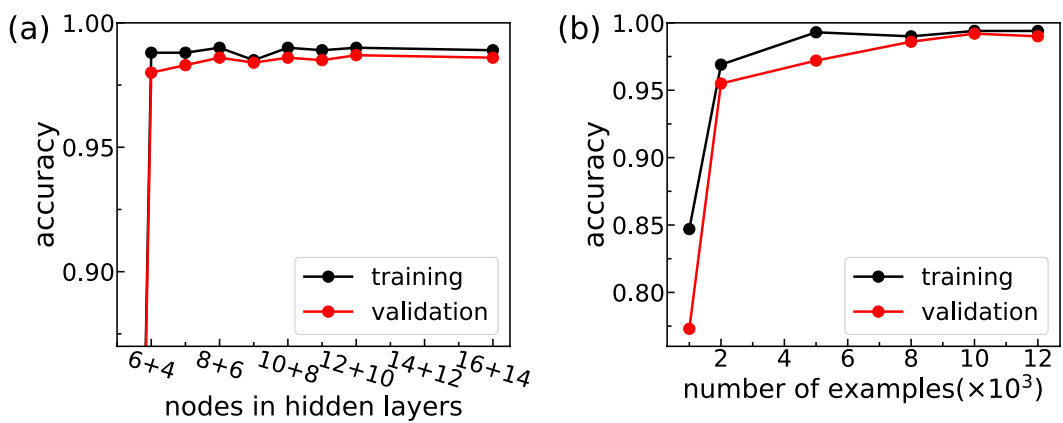

FIG. 3. Comparison of the prediction accuracies of the DNN classifier on the training and validation sets as a function of (a) number of nodes in the first and second hidden layers, (b) number of examples. The number of examples is defined as the sum of the number of training, validation and test sets, and training sets were composed of 60% of the total examples.

To quantitatively demonstrate how well the DNN classifier can generalize beyond the training dataset, Fig. 3 shows the performance of the network trained on different network architectures and various sizes of training sets, alongside the corresponding performance on the validation sets. The prediction accuracy, defined as the ratio of the number of correctly classified instances to the total number of instances, was adopted as a performance measure. In Fig. 3(a), the number of total examples is 8000 (4800 training, 1600 validation, and 1600 test). The mini-batch size in the gradient decent process was 32, together with a learning rate of 0.01. All the comparisons were based on the DNN trained over 1500 epochs, which was found to yield well-converged cross-entropy loss and reasonable generalization. The term "epoch" describes the process that all the training examples are fed into the DNN and all weights are updated accordingly in the backpropagation of the network. Since mini-batch gradient descent was applied, during each epoch, the weights were updated multiple times. The effect of the numbers of nodes in the two hidden layers, which are important indicators of the complexity of the DNN model, was first examined. A less complex model (fewer nodes in the hidden layers) is computationally easier to train but may suffer from underfitting; models with more nodes, on the other hand, generally fits the training sets well but may lead to poor generalization on the unseen data. Fig. 3(a) monitors the accuracies of the training and validation sets as a function of the number of nodes in the hidden layers. For every DNN model presented in Fig. 3(a), the second hidden layer has two fewer nodes than the first. The 6+4 configuration (6 nodes in the first hidden layer and 4 in the second) results in reasonably high accuracy on the training and validation sets, whereas simpler models, e.g., 5+3, yield large biases. The performance of the DNN seems to be converged with 8 and 6 nodes in the first and second hidden layers, respectively, and a more complicated model with more nodes does not noticeably increase the prediction accuracy on the validation sets. It was also found that the model performance is not strongly sensitive to the DNN architecture: the performance of the 16+14 network is comparable to that of 8+6, and no significant overfitting on the training set was identified. Therefore, the 8+6 is seemingly the optimal network configuration in terms of both prediction accuracy and model complexity. Fig. 3(b) demonstrates the performance of the 8+6 DNN trained on various number of examples. When the size of the entire dataset is less than 5000, the prediction accuracy on the training set is relatively low, and meanwhile the model exhibits high variance, i.e., the prediction accuracies on the validation sets are notably lower than that on the training sets. For comparison, the DNN trained on 8000 examples resulted in 98.9%, 98.6% and 98.7% accuracies on the training, validation and test sets, respectively, which indicates both high accuracy and low variance.

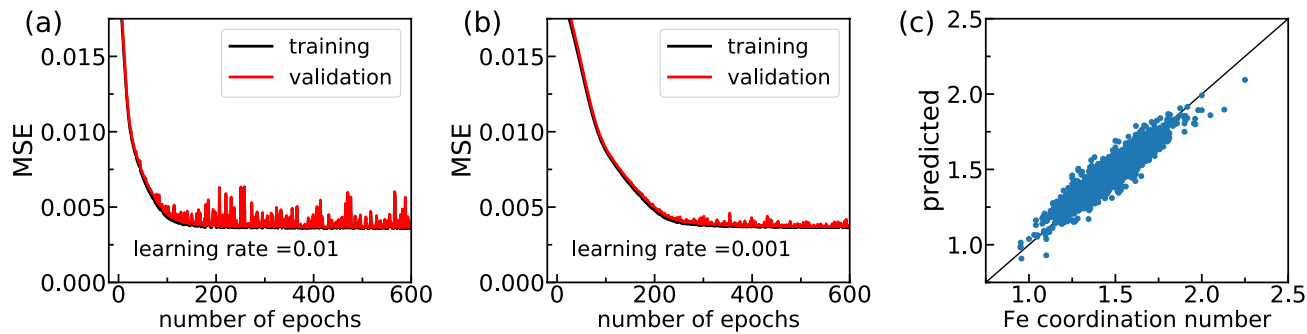

FIG.4. Performance of the DNN regressor on predicting the average Fe coordination numbers. MSE of the training and validation sets as a function of the number of epochs with a learning rate of (a) 0.01, (b) 0.001. (c) Comparison of the predicted and true Fe coordination numbers of the test set at the 400$^{th}$ epoch, when the model is trained with a 0.001 learning rate. The value of the MSE is 0.0035. The total number of examples used in the DNN regressor is 10000.

Fig. 4 demonstrates the performance of the DNN regressor on predicting the average Fe coordination numbers on an ensemble of oxygen atoms. The regressor architecture is similar to that of the classifier, with the main difference being that the output layer contains only one node and no activation function was applied on this node; this allows the regressor to generate continuous output values. The number of examples used in the training, validation and test sets are 6000, 2000 and 2000, respectively, and the mini-batch size used in training is 256. Fig. 4(a) shows the MSE values of the training and validation sets as a function of the number of epochs, with a learning rate of 0.01. The MSE values decrease rapidly and seem to converge after approximately 200 epochs. Since mini-batch gradient decent performs frequent weight updates and generally overshoots the global minimum of the loss function, the MSE values show large fluctuations as the training proceeds. This can be mitigated by using a smaller learning rate, as shown in Fig. 4(b). When the learning rate is decreased to 0.001, the MSE values show much smaller fluctuation but slower convergence compared with the results in Fig. 4(a), which was expected because of the reduced learning rate. Therefore, we chose 0.001 as the learning rate for the DNN regressor, and its performance was further examined on the test set. For a direct illustration, Fig. 4(c) plots the predicted vs. ground truth Fe coordination numbers of the test set after 400 epochs of training. It clearly shows a close agreement between the predicted and true values, with an MSE value of 0.0035, which correspond to a root-mean-squared error of 0.06, about 4% of the mean Fe coordination number.

In summary, this work proposed a simple DNN model composed of two hidden layers with 8 and 6 nodes, respectively, to perform structural prediction using XANES data. Depending on the prediction objective, the output layer of DNN can be adjusted to form a classifier or regressor to predict the local and averaged coordination environment, respectively. The effectiveness and accuracy of the model were demonstrated using the oxygen K-edge XANES of the $Li_3FeO_{3.5}$ system, and it is expected that this DNN model can also be utilized in other systems for structural predictions. In addition, both the DNN architecture and the input data can be enriched to perform more complicated prediction tasks. For instance, the input of the current model contains only continuous spectral data; to extend this model, categorial features of the materials, such as information on the space groups, absorbing atoms, or defect types, can also be incorporated into the input layer and converted into real-value vectors through embedding.[17] The DNN can then be expected to learn much more complicated features and be more versatile in predicting different material properties. Accordingly, more training examples collected on various absorbing atoms of a wide range of compounds are needed in order to build and train such a network.

**ACKNOWLEDGMENTS**

This work was supported as a part of the Center for Electrochemical Energy Science, an Energy Frontier Research Center funded by the US Department of Energy, Office of Science, Basic Energy Sciences under award number DE-AC02–06CH11. This work was performed, in part, at the Center for Nanoscale

Materials, an Office of Science user facility operated by Argonne National Laboratory, and supported by the U. S. Department of Energy, Office of Science, Office of Basic Energy Sciences, under Contract No. DE-AC02-06CH11357. The computing resources provided on Bebop, a high-performance computing cluster operated by the Laboratory Computing Resource Center at Argonne National Laboratory, is also gratefully acknowledged.